\documentclass[prc,twocolumn,a4paper,floatfix,nofootinbib,preprintnumbers,superscriptaddress,max]{revtex4-1}

\usepackage[utf8]{inputenc}
\usepackage{graphicx,psfrag}
\usepackage{mathrsfs}
\usepackage{amsmath,amsfonts,amssymb,booktabs}
\usepackage{multirow} 
\usepackage{diagbox}
\usepackage{comment}
\usepackage{xcolor}
\usepackage{enumerate}
\usepackage{orcidlink}
\usepackage{booktabs}
\usepackage{lineno}\usepackage{lipsum}
\usepackage{mathtools}
\usepackage{color,soul}
\usepackage{algorithm2e}
\usepackage{listings}
\usepackage{graphicx}
\usepackage{hhline}
\usepackage{float}
\usepackage{placeins}
\usepackage{hyperref}
\usepackage{cleveref}

\hypersetup{
    colorlinks = true,
    linkcolor = jsr,
    urlcolor = jsr,
    linkbordercolor = {white},
    citecolor = jsr,
    }

\usepackage{color}
\definecolor{cyan}{rgb}{0,0.9,0.9}
\definecolor{orange}{rgb}{0.9,0.5,0}
\definecolor{magenta}{rgb}{1,0,1}
\definecolor{purple}{rgb}{0.8,0.4,0.8}
\definecolor{gray}{rgb}{0.8242,0.8242,0.8242}
\definecolor{mgreen}{rgb}{0.1,0.8,0.1}
\definecolor{jsr}{HTML}{FF671F}

\usepackage[normalem]{ulem}

\def\bunine{\textsc{Bu2019lm}}
\def\katwo{\textsc{Ka2017}}
\def\possis{\texttt{POSSIS}}

\newcommand{\syserr}[1][]{\sigma_{\text{sys}#1}}

\newcommand{\xlan}{X_{\textrm{lan}}}
\begin{document}

\title{Data-driven approach for modeling the temporal and spectral evolution of kilonova systematic uncertainties}

\author{Sahil Jhawar \orcidlink{0000-0002-0003-0571}}
\email{jhawar@uni-potsdam.de}
\affiliation{Institut f\"ur Physik und Astronomie, Universit\"at Potsdam, Haus 28, Karl-Liebknecht-Str. 24/25, 14476, Potsdam, Germany}
\affiliation{GFZ German Research Centre For Geosciences, 14473 Potsdam, Germany}
\author{Thibeau Wouters \orcidlink{0009-0006-2797-3808}}
\affiliation{Institute for Gravitational and Subatomic Physics (GRASP), Utrecht University, Princetonplein 1, 3584 CC Utrecht, The Netherlands}
\affiliation{Nikhef, Science Park 105, 1098 XG Amsterdam, The Netherlands}
\author{Peter T. H. Pang \orcidlink{0000-0001-7041-3239}}
\affiliation{Nikhef, Science Park 105, 1098 XG Amsterdam, The Netherlands}
\affiliation{Institute for Gravitational and Subatomic Physics (GRASP), Utrecht University, Princetonplein 1, 3584 CC Utrecht, The Netherlands}
\author{Mattia \surname{Bulla} \orcidlink{0000-0002-8255-5127}}
\affiliation{Department of Physics and Earth Science, University of Ferrara, via Saragat 1, I-44122 Ferrara, Italy}
\affiliation{INFN, Sezione di Ferrara, via Saragat 1, I-44122 Ferrara, Italy}
\affiliation{INAF, Osservatorio Astronomico d’Abruzzo, via Mentore Maggini snc, 64100 Teramo, Italy
}
\author{Michael W. Coughlin \orcidlink{0000-0002-8262-2924}}
\affiliation{School of Physics and Astronomy, University of Minnesota, Minneapolis, Minnesota 55455, USA}
\author{Tim Dietrich \orcidlink{0000-0003-2374-307X}}
\affiliation{Institut f\"ur Physik und Astronomie, Universit\"at Potsdam, Haus 28, Karl-Liebknecht-Str. 24/25, 14476, Potsdam, Germany}
\affiliation{Max Planck Institute for Gravitational Physics (Albert Einstein Institute), Am M\"{u}hlenberg 1, Potsdam 14476, Germany}

\date{\today}

\begin{abstract}
    Kilonovae, possible electromagnetic counterparts to neutron star mergers, provide important information about high-energy transient phenomena and, in principle, also allow us to obtain information about the source properties responsible for powering the kilonova. 
    Unfortunately, numerous uncertainties exist in kilonova modeling that, at the current stage, hinder accurate predictions. Hence, one has to account for possible systematic modeling uncertainties when interpreting the observed transients. 
    In this work, we provide a data-driven approach to account for time-dependent and filter-dependent uncertainties in kilonova models. Through a suite of tests, we find that the most reliable recovery of the source parameters and description
    of the observational data can be obtained through a combination of kilonova models with time- and filter-dependent systematic uncertainties. We apply our new method to analyze AT2017gfo. While recovering a total ejecta mass consistent with previous studies, our approach gives insights into the temporal and spectral evolution of the systematic uncertainties of this kilonova. We consistently find a systematic error below $1$ mag between $1$ and $5$ days after the merger. Our work addresses the need for early follow-up of kilonovae at earlier times, and improved modeling of the kilonova at later times, to reduce the uncertainties outside of this time window.
\end{abstract}

\maketitle

\section{Introduction}
\label{sec:intro}
The gravitational-wave (GW) detection of the binary neutron star (BNS) merger GW170817~\cite{LIGOScientific:2017vwq,LIGOScientific:2017ync} by Advanced LIGO~\cite{LIGOScientific:2014pky} and Advanced Virgo~\cite{VIRGO:2014yos} GW observatories, combined with the observation of electromagnetic (EM) waves from the short gamma-ray burst GRB170817A and the kilonova AT2017gfo, has been a game changer in our understanding of, e.g, cosmology~\cite{LIGOScientific:2017adf,Guidorzi:2017ogy,Hotokezaka:2018dfi,Nakar:2020pyd,Coughlin:2019vtv,Kashyap:2019ypm,Coughlin:2020ozl,Dietrich:2020efo,Bulla:2022ppy,Gianfagna:2022kpw}, nuclear physics~\cite{Annala:2017llu,Bauswein:2017vtn,Fattoyev:2017jql,Capano:2019eae,LIGOScientific:2017vwq,LIGOScientific:2018hze,LIGOScientific:2018cki,Dietrich:2020efo,Huth:2021bsp}, modified theories of gravity~\cite{Ezquiaga:2017ekz,Baker:2017hug,Creminelli:2017sry}, and the chemical evolution of our Universe~\cite{Cowperthwaite:2017dyu,Smartt:2017fuw,Kasliwal:2017ngb,Kasen:2017sxr,Watson:2019xjv,Rosswog:2017sdn,Watson:2019xjv}. Most of these studies relied on the availability of information from multiple messengers, which in this case are, GWs and EM waves.

Although GW170817 has been the only multi-messenger detection of a BNS merger, there has been observational evidence that GW190425~\cite{LIGOScientific:2020aai}, GRB211211~\cite{Rastinejad:2022zbg,Ryan:2019fhz,Kunert:2023vqd} and GRB230307A~\cite{JWST:2023jqa} also originated from BNS mergers. Hence, the chances for further multi-messenger detections are continuously increasing due to the increasing range of observational facilities. For a reliable interpretation of GW170817 and future events, it is necessary to compare the observational data with theoretical predictions to extract characteristic information from the sources. The employed theoretical models must be accurate to ensure an unbiased estimate. Otherwise, the ever-increasing number of observational data or the observation of sources with higher accuracy would potentially lead to biased constraints.

Numerous studies have addressed the accuracy of GW models, e.g., Refs. \cite{Samajdar:2018dcx, Samajdar:2019ulq, Gamba:2020wgg, Kunert:2021hgm,Yelikar:2024rmh}, and it has been shown that systematic uncertainties are under control for the current generation of GW detectors; yet, better models are needed for the next generation of detectors, such as the Einstein Telescope~\cite{Punturo:2010zz,Sathyaprakash:2012jk, Branchesi:2023mws} and the Cosmic Explorer~\cite{Srivastava:2022slt}. In contrast, systematic uncertainties are more severe and not yet fully understood for kilonovae, an EM transient powered by the radioactive decay of unstable heavy elements synthesized through r-process nucleosynthesis ~\cite{Li:1998bw,Metzger:2019zeh} and spanning optical, infrared, and ultraviolet frequencies. Despite numerous efforts improving the modeling of kilonovae, e.g., by moving from simplified semi-analytical models (e.g, \cite{Grossman:2013lqa,Perego:2017wtu,Villar:2017wcc,Rosswog:2017sdn, Metzger:2019zeh,Barbieri:2019kli,Hotokezaka:2019uwo,Zhu:2020inc,Nicholl:2021rcr,Ricigliano:2023svx}) to more complicated full 3D radiative transfer simulations (e.g., \cite{Kasen:2013xka,Tanaka:2013ana,Wollaeger:2017ahm,Bulla:2019muo,Gillanders:2022opm,Collins:2022ocl}), there are still large uncertainties in the modeling. Among the main sources of uncertainty are properties of the ejected material \cite{Heinzel:2020qlt,Korobkin:2020spe,Tak:2023dfh,Fryer:2023osz,Tak:2024dlr} and some key ingredients setting the energy available to power the kilonova, i.e., nuclear heating rates and thermalization efficiencies, and the properties of the escaping radiation, i.e., the opacities of r-process elements \cite{Barnes:2020nfi,Zhu:2022qhc,Bulla:2022mwo,Brethauer:2024zxg,Sarin:2024tja,Wollaeger:2024xfx,Shenhar:2024rfm,Guttman:2024bxl}.

Given the significant uncertainties in kilonova modeling, it is essential to account for them during multi-messenger parameter inference. In the past, most works employing the Nuclear-physics and Multi-Messenger Astrophysics (NMMA) framework, which is the code infrastructure used for this article, accounted for these uncertainties by including a fixed systematic uncertainty of 1 magnitude~\cite{Dietrich:2020efo,Pang:2022rzc,Kann:2023ulv,Kunert:2023vqd}. This particular choice was motivated by the study of Ref.~\cite{Heinzel:2020qlt} where it was found that this uncertainty is sufficient to ensure that different ejecta morphologies assumed by different radiative transfer simulation codes make similar predictions, i.e., that the extracted ejecta properties are consistent within their statistical uncertainties. 
More recently, NMMA inference runs considered the uncertainty not as a fixed parameter, but as an additional sampling parameter~\cite{Hussenot-Desenonges:2023oll}.

Similar approaches have also been used in other works, e.g., \cite{Sarin:2023khf,Breschi:2024qlc,Peng:2024jqe}. In particular, Ref.~\cite{Sarin:2023khf} has updated its code infrastructure to account for such systematic uncertainty by adding a fixed systematic error standard deviation to the likelihood quadrature. A more flexible approach has been employed by Ref.~\cite{Peng:2024jqe} in which the systematic uncertainty is a sampling parameter, and Ref.~\cite{Breschi:2024qlc} allowed more flexibility to account for limitations in the employed kilonova models by adding different systematic uncertainty priors across different observational bands.

However, the systematic uncertainty of a kilonova model will, in general, not only depend on the observational bands, i.e., filter-dependent, but will also vary over time. This time dependence appears naturally when accounting for known uncertainties in the description of opacities at early times \cite[e.g.][]{Tanaka:2019iqp,Bulla:2022mwo} or non-local thermal equilibrium effects that become important after about 1 week \cite{Pognan:2022pix,Pognan:2021wpy}. 
 
In this work, we introduce a data-driven scheme for handling the time- and \textcolor{red}{filter}-dependent uncertainties in the light curve models. This method promises to improve the robustness of the parameter estimations. It is model-agnostic and can be applied to any EM model irrespective of the nature of the transient, e.g., also for gamma-ray-burst afterglows or supernovae.

This paper is structured as follows. Section~\ref{sec:methods} describes the methodological approach, including the radiative transfer models, the Bayesian inference framework, and the systematic error interpolation schemes employed. Section~\ref{sec:validation} provides validation tests using synthetic light curve data from different models. 
Section~\ref{sec:analysis} demonstrates the application of the methodology to AT2017gfo. Finally, Section~\ref{sec:conclusion} summarizes the key findings and outlines future perspectives. 

\section{Methods}
\label{sec:methods}

\subsection{Kilonova Light curve Computation}
An accurate description of the observables of kilonovae, i.e., luminosities, spectra, light curves, and polarization, requires detailed modeling of the radiation processes, incorporating the interaction between the radiation and the matter via absorption and scattering processes. In the following, we are employing results from two independent radiative-transfer codes to enable cross-validation and testing of our approach.

\possis~\cite{Bulla:2019muo,Bulla:2022mwo} and \texttt{SEDONA} \cite{Kasen:2006ce} are 3D Monte Carlo radiative transfer codes that model synthetic observable (flux and polarization spectral \textcolor{red}{times} series) for explosive transients such as supernovae and kilonovae. Both codes incorporate time-dependent opacities and ejecta properties, enabling them to capture the evolving nature of astrophysical events over various time scales. Inside the model grid, each cell is represented by velocity $v$, time-dependent density $\rho(t)$, time-dependent temperature $T(t)$, and the electron fraction $Y_e(t)$, starting at some reference time $t_0$.

The computation of the observables is based on the simulation of Monte Carlo photon packets diffusing through the freely expanding medium. Monte Carlo photons are created with energy and frequencies set by the specific emission process (e.g.\ radioactivity) and are propagated according to the opacity of the expanding medium. Those that escape the medium are then used to construct spectral-time series (from ultraviolet to infrared) at different viewing angles, from which light curves in different filters can be constructed.

\subsubsection{\bunine}
\label{subsec:possis}

While \possis\ can generally support arbitrary geometries, we focus here on the usage of a two-component model grid; cf.\ Ref.~\cite{Dietrich:2020efo} for more details. The first component characterizes dynamical ejecta and contains a lanthanide-rich part around the equatorial plane and a lanthanide-free part at the polar regions. The second component accounts for wind-driven ejecta and is spherically symmetric. The model, hereafter denoted with \bunine, is parameterized by the ejecta mass of the two components, $M_{\rm ej,dyn}$ and $M_{\rm ej,wind}$, the half-opening angle of the lanthanide-rich dynamical-ejecta component, $\phi$, and the viewing angle $\theta_{\rm obs}$.

\subsubsection{\katwo}
\label{subsec:kasen}

Although \texttt{SEDONA} is a 3D code, here we focus on a grid of one-dimensional spherically symmetric models presented in Ref.~\cite{Kasen:2017sxr}. The model, in the following referred to as \katwo, is parameterized by the ejecta total mass $M_{\rm ej}$, average velocity $v_{\rm ej}$, and lanthanide fraction\ $\xlan$. The bulk of the freely expanding ejecta is determined by the ejecta mass $M_{\rm ej}$. The density profile of the ejecta is described using a broken power-law that transits from the gradually declining interior with $v_{\rm ej}t / r$ to the steeply dropping outer layer with $(v_{\rm ej}t / r)^{10}$. Finally, the lanthanide fraction $\xlan$ influences the opacity and color evolution of the kilonova, where larger lanthanide fractions result in increased opacity and longer-duration emissions shifted towards the infrared.

\subsection{Surrogate kilonova models}

The \possis\ and the \texttt{SEDONA} codes are computationally too expensive to be run on the fly during sampling. Therefore, we train surrogate models for the \bunine\ and \katwo\ grids that are cheaper to execute during Bayesian inference. 

For \bunine, we use the dataset of $1596$ parameter combinations and their light curves generated by \possis\ to create the surrogate model. For each filter that we consider in our analyses below, we perform a singular-value decomposition (SVD) to reduce the dimensionality of the output, setting the number of SVD components to $10$. Then, we train a fully connected artificial neural network that maps the values of the \bunine\ parameters to the SVD coefficients, from which the light curve can be reconstructed. The architecture of the neural network consists of three hidden layers, having $128$, $256$, and $128$ neurons, respectively. Training is done with \texttt{TensorFlow}~\cite{tensorflow2015-whitepaper} and runs with the Adam optimizer~\cite{Kingma:2014vow}, with a fixed learning rate of $10^{-3}$ and a batch size of $128$, for $100$ epochs. We rescale the input and output data with a min-max scaler before training and use $20\%$ of the dataset as validation data to ensure that the network is not overfitting.

For \katwo, we use the publicly-available\footnote{\href{https://github.com/dnkasen/Kasen_Kilonova_Models_2017}{https://github.com/dnkasen/Kasen\_Kilonova\_Models\_2017}} kilonova light curves produced with \texttt{SEDONA}. The full dataset contains 329 parameter combinations, which are used to create the surrogate model. A similar SVD and neural network training is performed for the \katwo\ model as for \bunine. However, here we we used a hidden layer of $2048$ neurons, followed by a dropout layer with a dropout rate of $0.6$.

To quantify the performance of our surrogate models, we compute the Root Mean Square Error (RMSE). The RMSE quantifies the average deviation between model predictions and actual values, which is defined as
\begin{equation}
    \text{RMSE} = \sqrt{\frac{1}{N} \sum_{i=1}^N\left(y_i-\hat{y}_i\right)^2}\ ,
\end{equation}
where $y_i$ is the actual value obtained through the radiative-transfer simulation at time $t_i$, $\hat{y}_i$ denotes the prediction from the surrogate model, and $N$ is the total number of time points. We compute this metric across our entire parameter grid and all filters and report its median value.

We find median RMSE values of $0.103~\rm mag$ for \bunine\ (computed over a 14-day period) and $0.485~\rm mag$ for \katwo\ (computed over a 7-day period).

As discussed later in Sec.~\ref{ssec:time_filter_uncer} regarding time- and filter-dependent uncertainties, the median $\syserr$ is $0.291~\rm mag$, greater than the surrogate model's median RMSE. This suggests that the \bunine\ surrogate model is sufficiently accurate for our study, and we are not over-confident about it.
\subsection{Bayesian Inference}\label{sec:bayesian_inference}

By using Bayes' theorem, the posterior $p(\vec{\theta} | d,\mathcal{H})$ on a set of parameters $\vec{\theta}$ under the hypothesis $\mathcal{H}$ and with data $d$ is given by
\begin{equation}
		p(\vec{\theta} | d,\mathcal{H}) =  \frac{p(d|\vec{\theta},\mathcal{H})p(\vec{\theta}|\mathcal{H})}{p(d|\mathcal{H})},
\end{equation}
or in short form: 
\begin{equation}
\mathcal{P}(\vec{\theta}) =  \frac{\mathcal{L}(\vec{\theta})\pi(\vec{\theta})}{\mathcal{Z}}\,,
\end{equation}
where $\mathcal{P}(\vec{\theta})$, $\mathcal{L}(\vec{\theta})$, $\pi(\vec{\theta})$, and $\mathcal{Z}$ are the posterior, likelihood, prior, and evidence, respectively.
The prior describes our knowledge of the parameters before any observations. The likelihood quantifies how well the hypothesis can describe the data at a given point $\vec{\theta}$ in the parameter space.
Finally, the evidence, also known as the marginalized likelihood, marginalizes the likelihood over the whole parameter space with respect to the prior, i.e.,
\begin{equation}
	\mathcal{Z} = \int d\vec{\theta} \mathcal{L}(\vec{\theta})\pi(\vec{\theta})\,.
\end{equation}



\begin{figure}
	\begin{center}
		\includegraphics[width=0.5\textwidth]{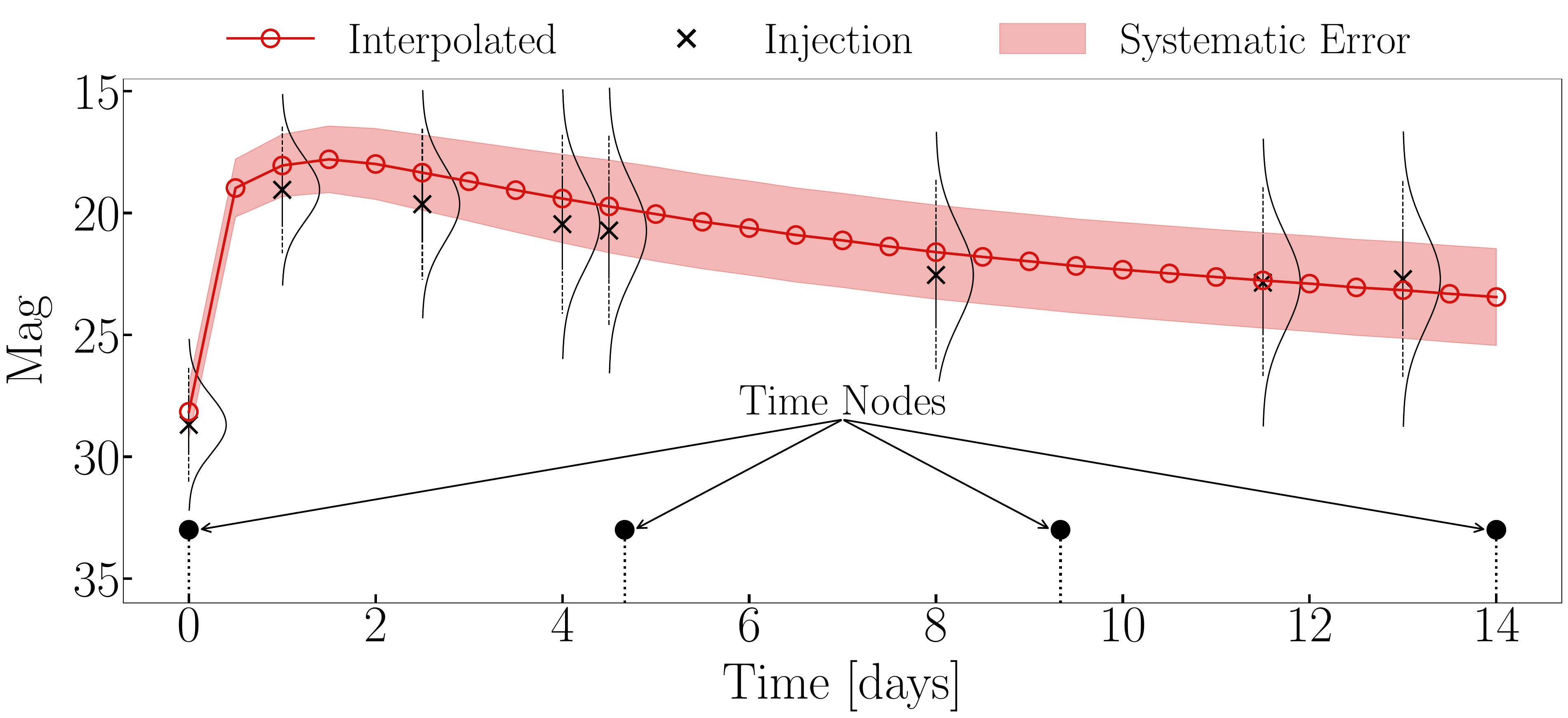}
		\caption{The stem plot shows the placement of four time nodes at 0, 4.67, 9.34, and 14 days used for interpolation.}
		\label{fig:sys_err_combined}
	\end{center}
\end{figure}

\subsection{Time-dependent systematic error}\label{ssec:time} 

To parameterize the time-dependence of the systematic uncertainty $\sigma(t)$, we make use of a piecewise linear interpolation scheme with $N$ evenly spaced time nodes
\begin{equation}
\begin{aligned}
	\label{equation:linear_interp}
	\sigma_{\rm sys}(t) &= \sigma_n+\frac{\sigma_{n+1}-\sigma_n}{t_{n+1}-t_n} \cdot (t-t_n), \ {\rm for \ } t_n \leq t < t_{n+1} \\
 &{\rm with} \ n = 0, 1, ..., N - 1.
 \end{aligned}
\end{equation}
Each $\sigma_n$ has a uniform prior in $[a, b)$, where $a$ and $b$ are the assumed lower and upper bound of the systematic uncertainty. Throughout our analyses, we used a prior of $\mathcal{U}(0,2)$.

The likelihood function $\mathcal{L}(\vec{\theta})$ is given by
\begin{equation}
	\label{equation:likelihood}
	\mathcal{L}(\vec{\theta}) \propto \exp \left(-\frac{1}{2} \sum_{i j} \frac{\left(m_i^j-m_i^{j, \text { est }}(\vec{\theta})\right)^2}{\left(\sigma_i^j\right)^2+\left(\syserr[,i]\right)^2}\right),
\end{equation}
where $m^j_i$ is the AB magnitude in filter $j$ at time $i$ with the corresponding measurement error $\sigma^j_i \equiv \sigma^j(t_i)$, $m_i^{j, \text { est }}(\vec{\theta})$ is the estimated AB magnitude for the source parameters $\vec{\theta}$ (e.g., ejecta masses, velocities) from the model and $\syserr[,i]$ is the interpolated systematic error at time $t_i$.

The described procedure is sketched in Fig.~\ref{fig:sys_err_combined}; where black crosses mark the mock data. The red line represents the best-fit light curve, the red-shaded region represents the systematic uncertainty in the best-fit posterior, and the Gaussian curve around them represents the denominator of~\Cref{equation:likelihood}.

\subsection{Time- and filter-dependent systematic error}\label{ssec:filter}

To extend our systematic error analysis, we also implement an additional filter-dependence. This enables us to consider more general cases in which uncertainties vary in time, and the accuracy of model predictions is filter dependent, e.g., due to filter-dependent uncertainties of the opacities; cf. Ref.~\cite{Tanaka:2019iqp}.
For this purpose, we will allow different systematic errors, $\syserr[]^j(t_i)$ for different observational filters. In this approach, the filters that need to be sampled independently and jointly are based on evaluating the Mean Absolute Deviation (MAD) for each filter and comparing it to the overall MAD calculated across all filters.

The MAD is a statistical measure of the variability and dispersion of data values, and it is used here to determine the extent to which each filter contributes to the overall variability of the data.
The MAD for each filter $j$ is calculated as
\begin{equation}
	\label{equation:mad_filt}
	\mathrm{MAD}_{j} = \frac{1}{n_{j}}\sum_{i=1}^{n_j} |x_{i,j} - \mu_{j}|,
\end{equation}
where $x_{i,j}$ is the AB magnitude in filter $j$ at time $t_i$, $\mu_{j}$ is the mean of all AB magnitudes in filter $j$, and $n_j$ is the total number of data points in filter $j$. Similarly, the total MAD is calculated as
\begin{equation}
	\label{equation:mad_total}
	\mathrm{MAD}= \frac{1}{n}\sum_{i=1}^n|x_{i} - \mu|,
\end{equation}
where $x_{i}$ is the AB magnitude at time $t_i$, $\mu$ is the mean of all AB magnitudes and $n$ is the total number of data points across all available filters.

The choice of filter for independent and joint systematic error is based on the comparison of the filter-specific $\mathrm{MAD}_{j}$ to $\mathrm{MAD}$.
Given this, the likelihood in~\Cref{equation:likelihood} can be re-written as

\begin{equation}
	\label{equation:likelihood_f}
	\mathcal{L}(\vec{\theta}) \propto \exp \left(-\frac{1}{2} \sum_{i j} \frac{\left(m_i^j-m_i^{j, \text { est }}(\vec{\theta})\right)^2}{\left(\sigma_i^j\right)^2+\left(\syserr[,i]^j\right)^2}\right),
\end{equation}
where $\syserr[,i]^j$ is the interpolated systematic error at time $t_i$ and filter $j$.

Such a likelihood is equivalent to including an additional shift to the light curve by $\Delta m$, and marginalizing it with a normal distribution with a mean of 0 and variance of $\syserr[,i]^j$.

\begin{table}[t]
    \centering
    \renewcommand{\arraystretch}{1.35}
    \begin{tabular}{l|cc}
        \hline\hline
        \diagbox[width=2.7cm, innerrightsep=10pt]{Parameter}{Model} & \bunine  & \katwo  \\\hline
        $D_L [\rm{Mpc}]$                        & 40 & 40\\
        $\log_{10}M_{\rm{ej}} [M_\odot]$        & \multirow{3}{*}{--} & $-1.43$       \\
        $\log_{10}v_{\rm{ej}} [\rm{c}]$         &                     & $-0.74$      \\
        $\log_{10}X_{\rm{lan}}$                 &                     & $-3.38$     \\
        $\Phi [\rm{deg}]$                       & 68.69 & \multirow{4}{*}{--}\\
        $\iota [\rm{rad}]$                      & 0.43  & \\
        $\log_{10}M^{\rm{dyn}}_{\rm{ej}} [M_\odot]$  & $-1.18$ & \\
        $\log_{10}M^{\rm{wind}}_{\rm{ej}} [M_\odot]$ & $-2.25$ & \\ \hline\hline
    \end{tabular}
    \caption{Parameter values used to generate the mock light curves to test the implemented algorithm.}
    \label{tab:injection}
\end{table}

\begin{figure*}[t]
	\begin{center}
		\includegraphics[width=0.9\textwidth]{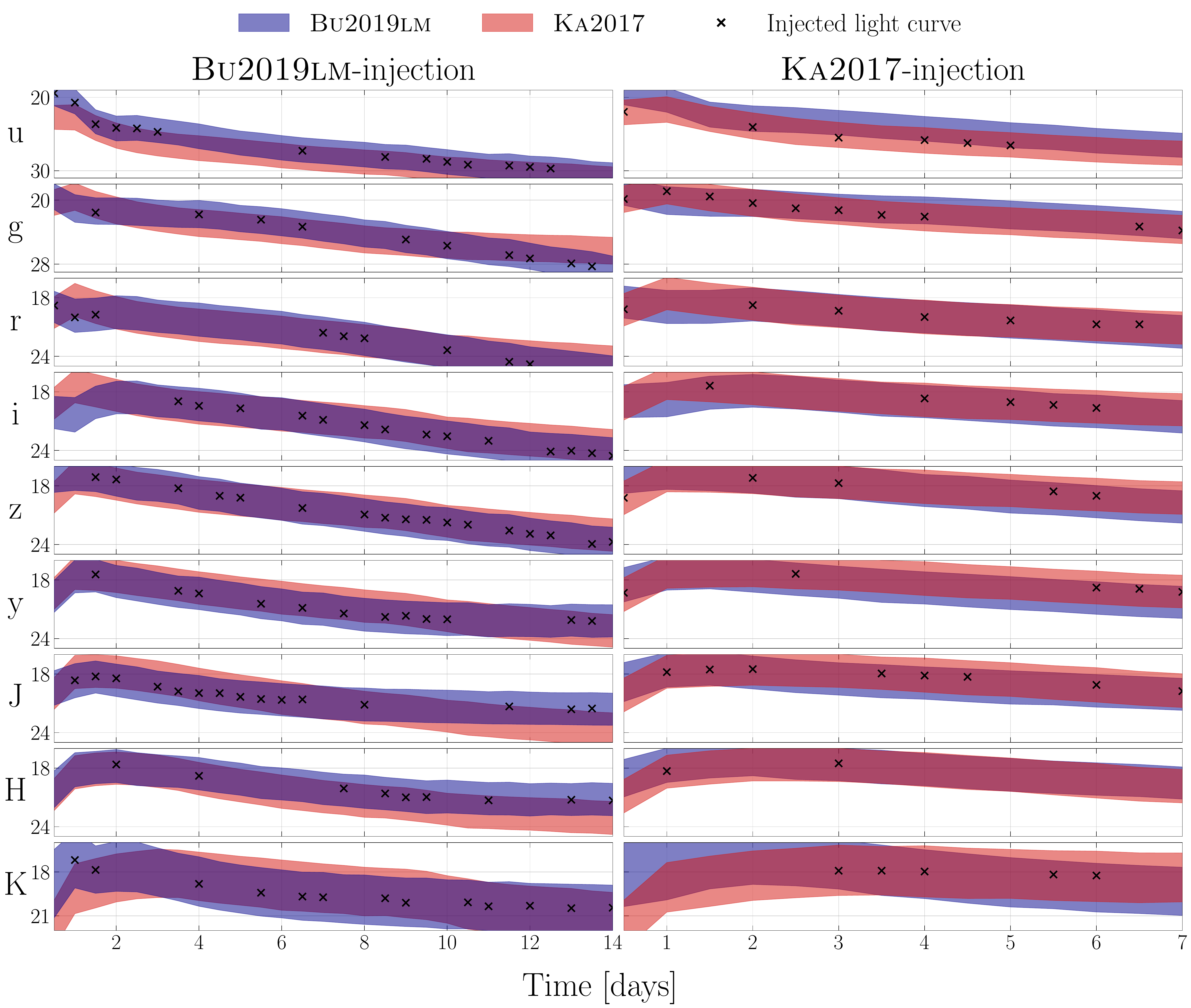}\\
        \includegraphics[width=0.45\textwidth]{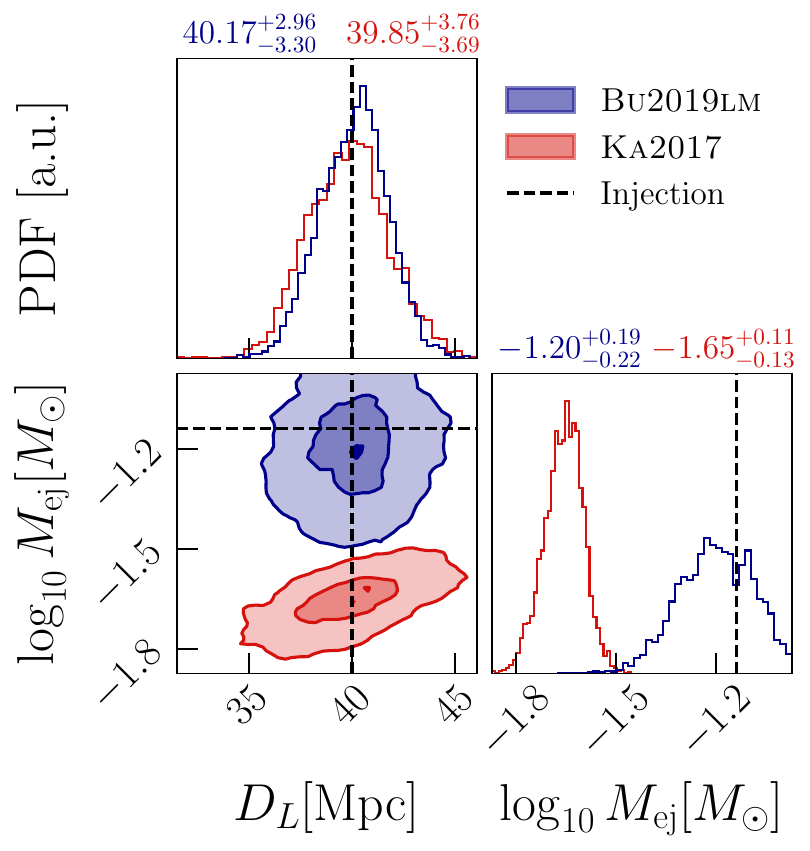}
  	\includegraphics[width=0.45\textwidth]{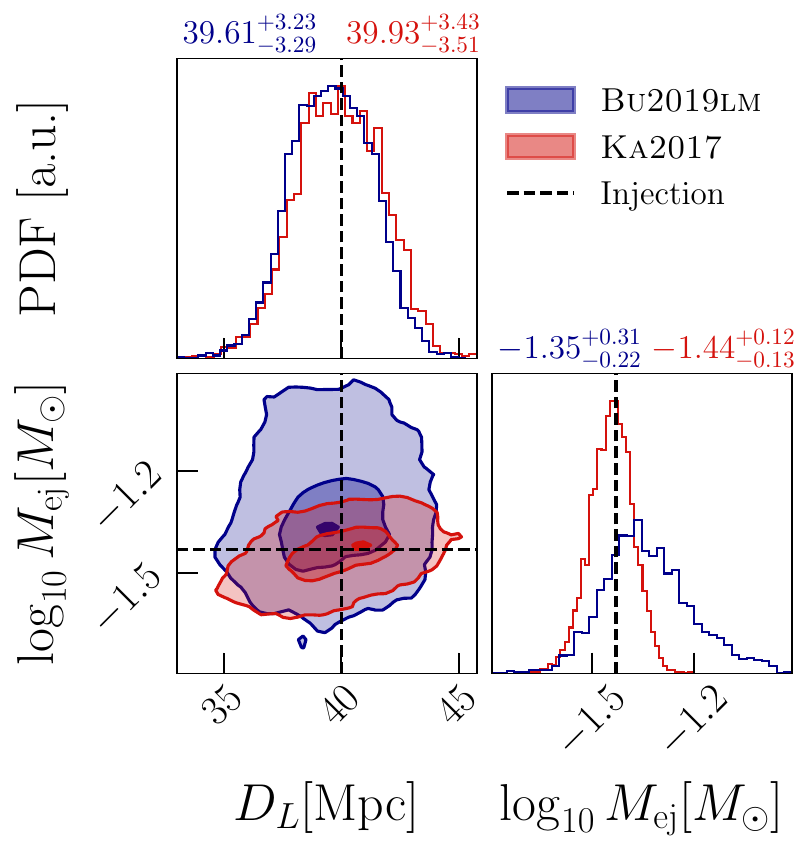} 
		\caption{\textit{Top panels:} Light curves for the validation test employing a constant 1-magnitude uncertainty. The crosses represent the injected light curve employed as input for the Bayesian inference. The band represents the 90\% credibility region of the light curves generated from the posterior samples.
        \textit{Bottom panels:} 2D marginalised posteriors of \textsc{Bu2019lm} (left) and \textsc{Ka2017} (right) with $2\sigma$ shaded region and injected parameter.}
        \label{fig:Ka2017_Bu2019lm_injection_1mag}
	\end{center}
\end{figure*}

\begin{figure*}
	\begin{center}
		\includegraphics[width=0.9\textwidth]{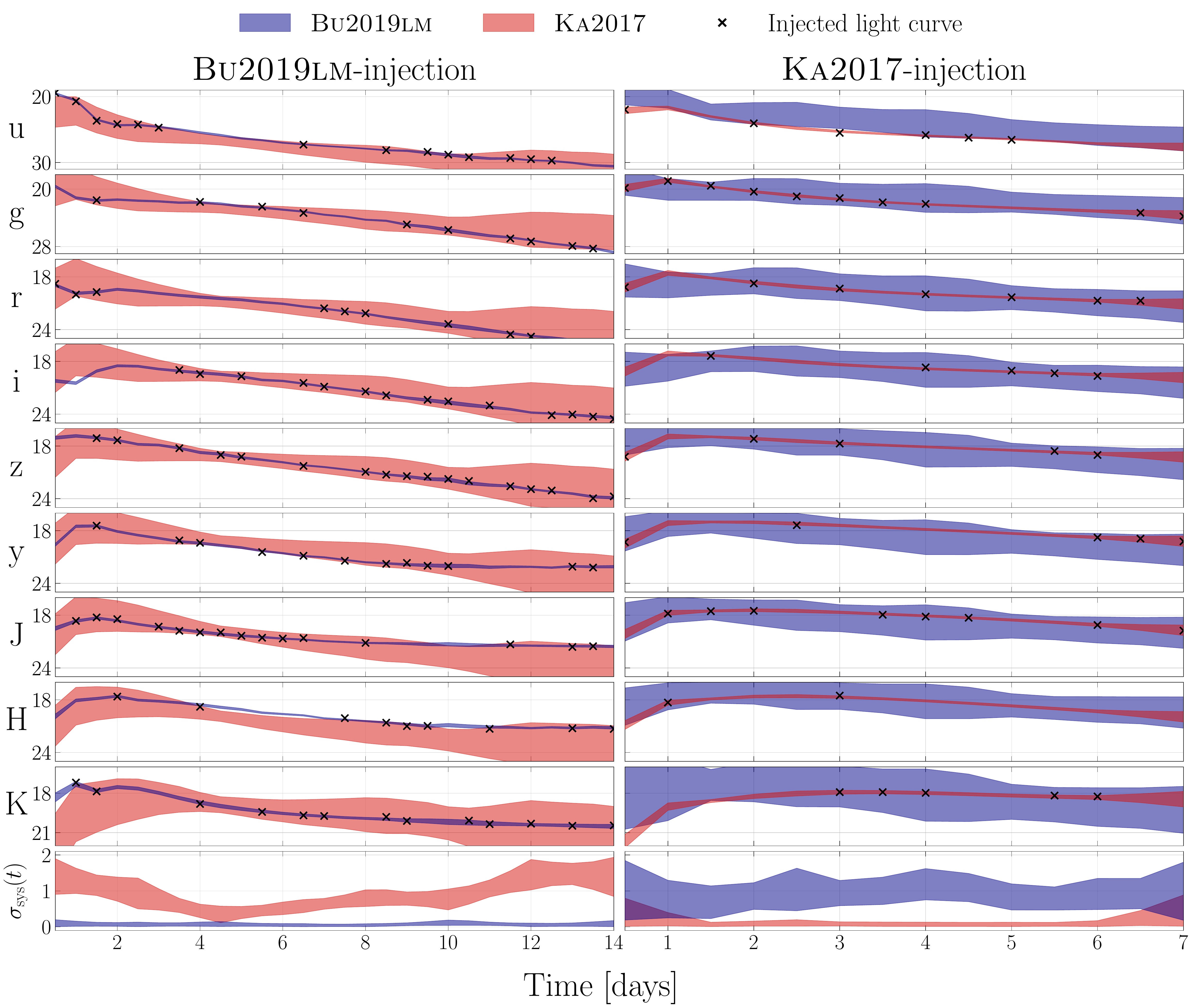}\\
        \includegraphics[width=0.45\textwidth]{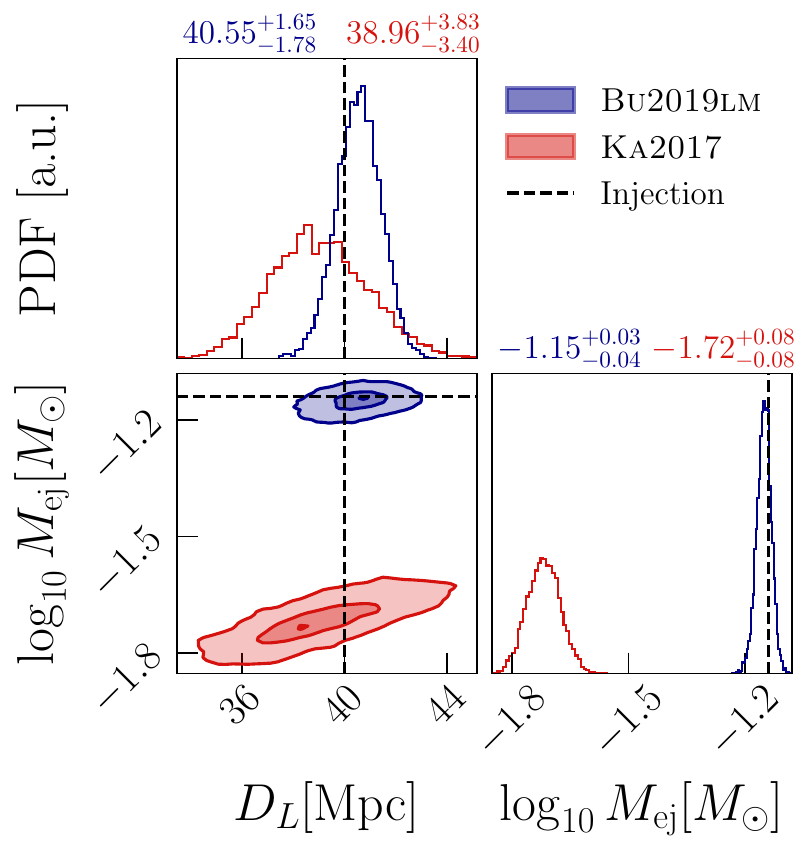}
  	\includegraphics[width=0.45\textwidth]{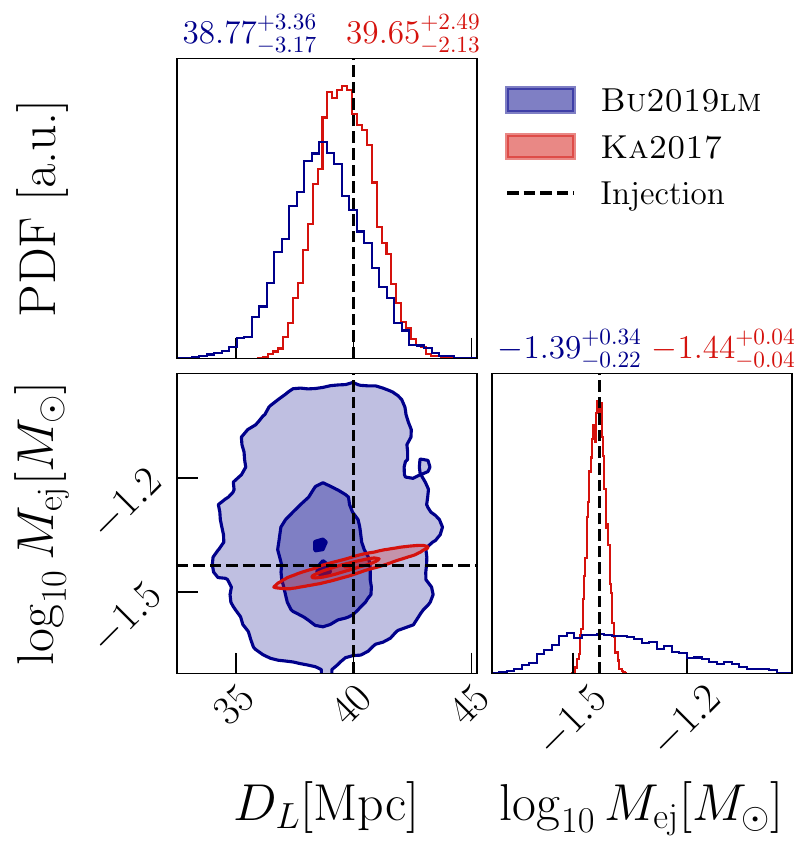} 
	
		\caption{Same as Fig.~\ref{fig:Ka2017_Bu2019lm_injection_1mag}, for time-dependent systematic uncertainty. The bottom panel in the light curve plots illustrates the time-discretized systematic uncertainty, where the band represents the 90\% credibility region of the re-interpolated systematic uncertainty, $\syserr(t)$.}
        \label{fig:Ka2017_Bu2019lm_injection_timedependent}
	\end{center}
\end{figure*}

\begin{figure*}[t]
	\begin{center}
		\includegraphics[width=0.9\textwidth]{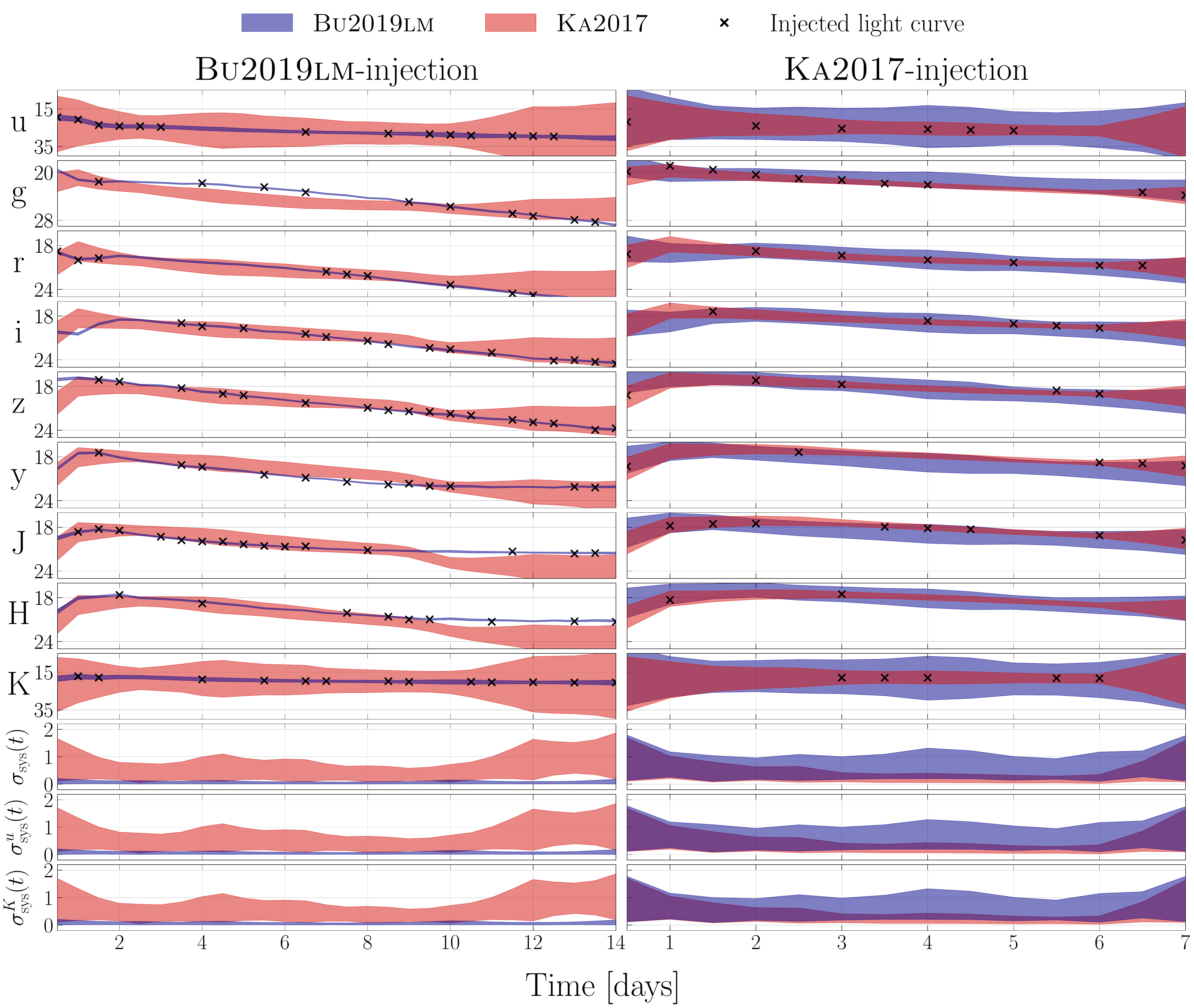}\\
        \includegraphics[width=0.45\textwidth]{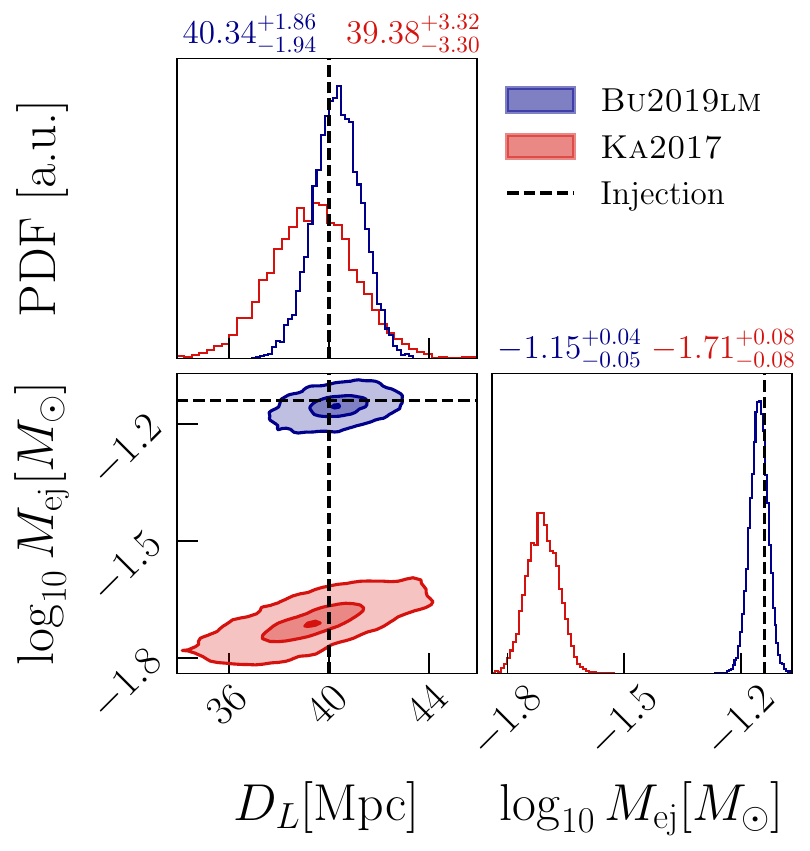}
  	\includegraphics[width=0.45\textwidth]{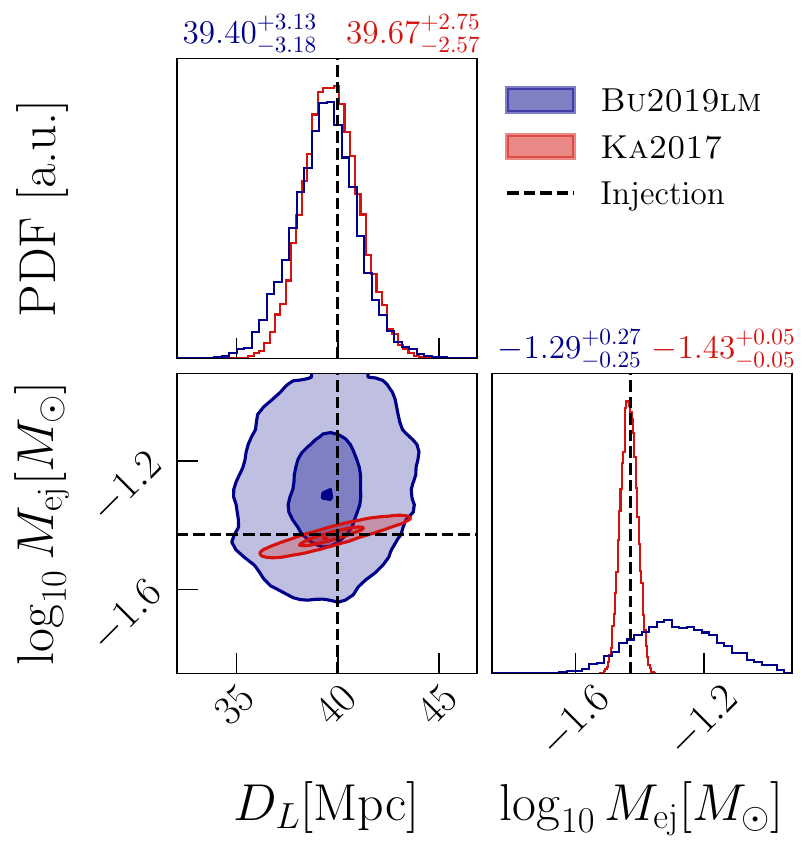} 
		\caption{Same as Fig.~\ref{fig:Ka2017_Bu2019lm_injection_timedependent}, for time- and filter-dependent systematic uncertainty. $\syserr[]^u(t)$ and $\syserr[]^K(t)$ represents systematic uncertainty for independently sampled $u$ and $K$ band, $\syserr(t)$ is for rest all the bands sampled together.}
        \label{fig:Ka2017_Bu2019lm_injection_uK}
	\end{center}
\end{figure*}

\section{Validation}
\label{sec:validation}

To validate our methodology, we simulated two synthetic light curves employing the \bunine\ and \katwo\  models. 
We use a uniform time step of $0.5$ days for sampling and randomly select 45\% of the data (until $20$ days) to account for partially missing data due to the `lack' of observations, e.g., through bad weather conditions or other observational limitations. To account for errors in the observations, we add a random shift to each datapoint following a Gaussian distribution with zero mean and a standard deviation of ${0.1 ~\rm mag}$.

\FloatBarrier

These simulated data serve as \textit{injections} and are taken for up to one week for \katwo\ and upto two weeks for \bunine\ after the merger, i.e., later data is not injected into the parameter estimation pipeline.\footnote{We decided to reduce the length of the injection of the \katwo\ model to 1 week since some of the lightcurves show unphysical features after about 1 week, e.g., an increasing luminosity due to low signal-to-noise from the underlying simulations at late epochs. However, we have also checked our results to be robust for a length of 10 and 14 days.}

For each injection, we use both models (\bunine\ and \katwo) for the recovery. The employed injection parameters are inspired by the parameters of AT2017gfo and summarized in Tab.~\ref{tab:injection}. To compare the posterior distributions of two competing models, we will mainly focus on the common parameters, i.e., the luminosity distance, $D_L$, and the total ejecta mass, $\log_{10}M_{\rm{ej}}$, where,
$M_{\textrm{ej}} =  {M_{\mathrm{ej,wind}}} +{M_{\mathrm{ej,dyn}}}$ for the \bunine\ model.

\subsection{Constant Systematic Uncertainties}

We start our analysis following a similar approach as employed in our previous studies, e.g.,~\cite{Dietrich:2020efo,Heinzel:2020qlt,Pang:2022rzc}, using a constant 1\,mag uncertainty. The obtained results are summarized in Fig.~\ref{fig:Ka2017_Bu2019lm_injection_1mag}, where one can see that for all injections, the assumed 1\,mag uncertainty is sufficient for the recovered light curves to approximate the injected light curves reliably, even when different models are used for the injection and recovery.

Considering the recovery of the injected parameters, we find that when using the same model for the injection and the recovery, the injected parameters can be recovered reliably within the statistical uncertainties without any visible bias. 
However, when a different model is employed for the recovery, we find that there can be systematic biases in the recovered posteriors of the source parameters. Generally, the \bunine\ model recovers injected values for both models better than the \katwo\ model, which we assume is due to the two-component ejecta, which enables more flexibility during the light curve fitting.

\subsection{Time-dependent Uncertainties}

Relaxing the assumption of a constant 1\,mag uncertainty and enabling a time-dependent uncertainty, we find significantly different results regarding the accuracy of the recovered light curves. 
In fact, using the same model for the injection and recovery can be considered our best-case scenario, in which the model is completely accurate and accounts for all the relevant physics of the kilonova light curve with respect to the model used for injection and recovery. Therefore, one can expect the estimated systematic uncertainty $\sigma_{\rm sys}(t)$ to be minimal across the whole time range. This expectation is fully confirmed by our test and is visible in Fig.~\ref{fig:Ka2017_Bu2019lm_injection_timedependent}, where the 90\% posterior light curve band is extremely tight around the injected data and the injected light curve falls within the band across all filters and times. The obtained uncertainty of $\mathcal{O}(0.1\rm mag)$ is dominated by the uncertainty added to the injection data mimicking the uncertainties in obtained observational data. Varying these uncertainties, we verified that our method is able to pick up larger uncertainties if there is a larger spread in the injected data points. 

This improved recovery of the light curves also leads to posteriors recovering the injected value with a smaller spread around the injected value; being more quantitative, we find a reduction of the spread of the posterior by up to a factor of two. This is clearly visible in Fig.~\ref{fig:Ka2017_Bu2019lm_injection_timedependent}.

As illustrated in Fig.~\ref{fig:Ka2017_Bu2019lm_injection_timedependent}, when injecting \katwo\ light curves, both models successfully recover the injected parameters. However, the \bunine\ model's posterior is significantly broader (with more than 5 times larger uncertainty), which clearly indicates that with sufficient flexibility in systematic errors, \bunine\ can achieve satisfactory performance. In contrast, when attempting to recover \bunine\ injections with the \katwo\ model, we observe a significant discrepancy. The recovered mass deviates substantially from the combined injected masses, suggesting that the \katwo\ model, being a spherically symmetric, single component ejecta model, is too limited to accurately represent kilonova light curves that have a larger variability and complexity.

\subsection{Filter-dependent Uncertainties}

Finally, we present in Fig.~\ref{fig:Ka2017_Bu2019lm_injection_uK} an analysis in which we employ a time- and filter-dependent uncertainty during our recovery. Based on initial tests, in particular when studying AT2017gfo, we have found the most significant differences in the ultraviolet $u$-band and the infrared $K$-band. For this reason, we decided to allow different systematic uncertainties in these two bands and group all other bands, i.e., $g$ to $H$, together using the same uncertainty. Clearly, this particular choice is not unique, and numerous other options would be possible, e.g., employing different uncertainties for all filters. However, even when grouping the bands as described above, our time- and filter-dependent uncertainty analysis has three times as many free uncertainty parameters as the analysis shown in the previous subsection, which uses purely time-dependent uncertainties, which increases the runtime of the analysis. 
As in the previous case, we find that using the same model for recovery as for the injections results in an accurate description of the light curve and a recovery of the source parameters with small uncertainties. The recovery is similar to the time-dependent results. 
Similarly, when the recovery is based on a different model than the one used for creating the injection data, we find -as before- that, in particular, the \katwo\ model fails in recovering the correct injection mass of the \bunine\ injection. 

\begin{figure}
	\begin{center}
		\includegraphics[width=0.5\textwidth]{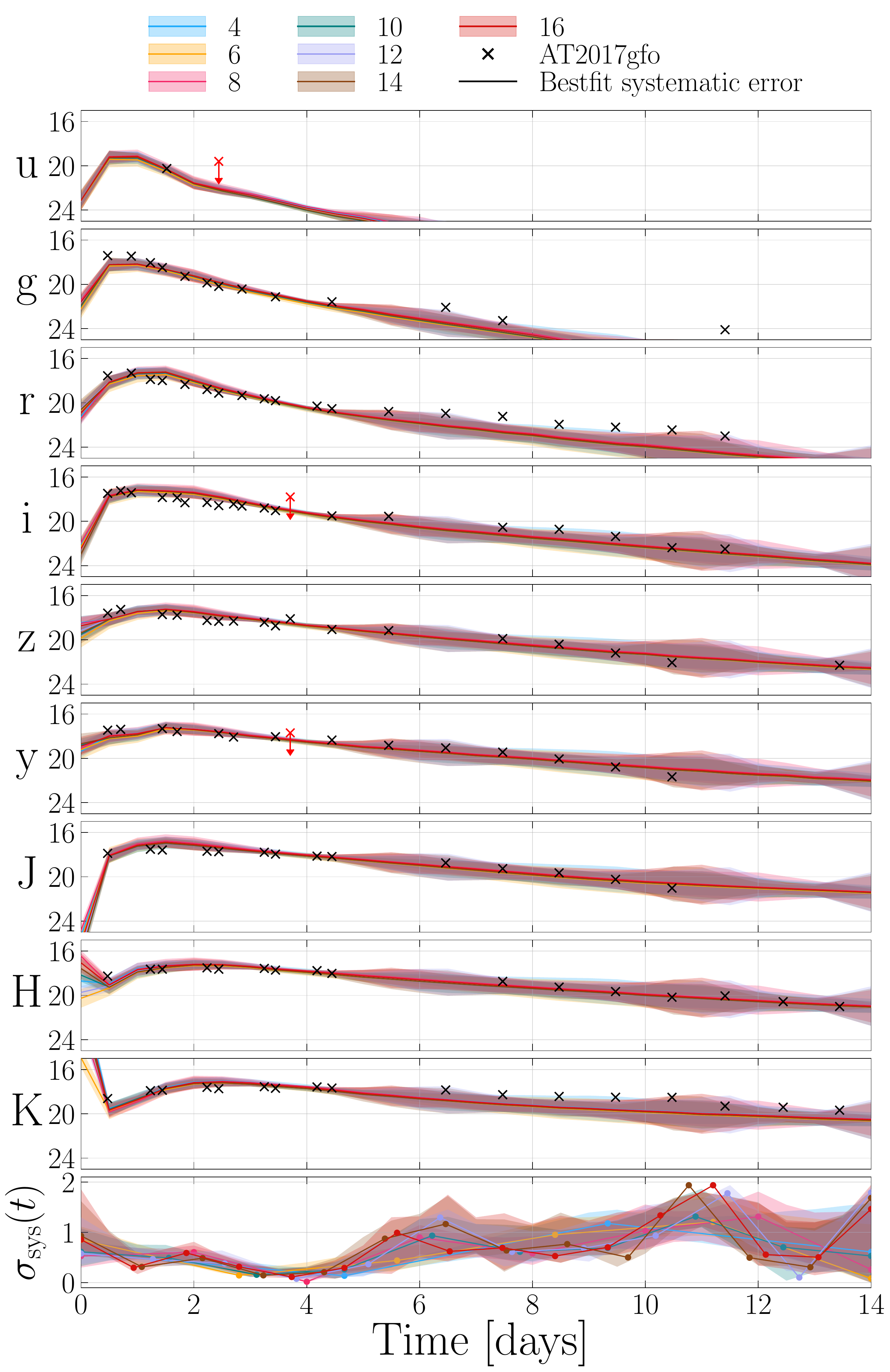}
		\caption{Recovered light curves for AT2017gfo in different observational bands employing the time-dependent uncertainty method described in the main text. Different colors represent different numbers of employed time nodes. Observational data are marked with black crosses, while red crosses mark the non-detections. The bottom panel illustrates the time-discretized systematic uncertainty, where the circle represents the systematic uncertainty corresponding to the placement of time nodes, and the band represents the 90\% highest density interval of the re-interpolated systematic uncertainty, $\syserr(t)$.}
		\label{fig:at2017gfo}
	\end{center}
\end{figure}

\begin{figure}
	\begin{center}
		\includegraphics[width=0.5\textwidth]{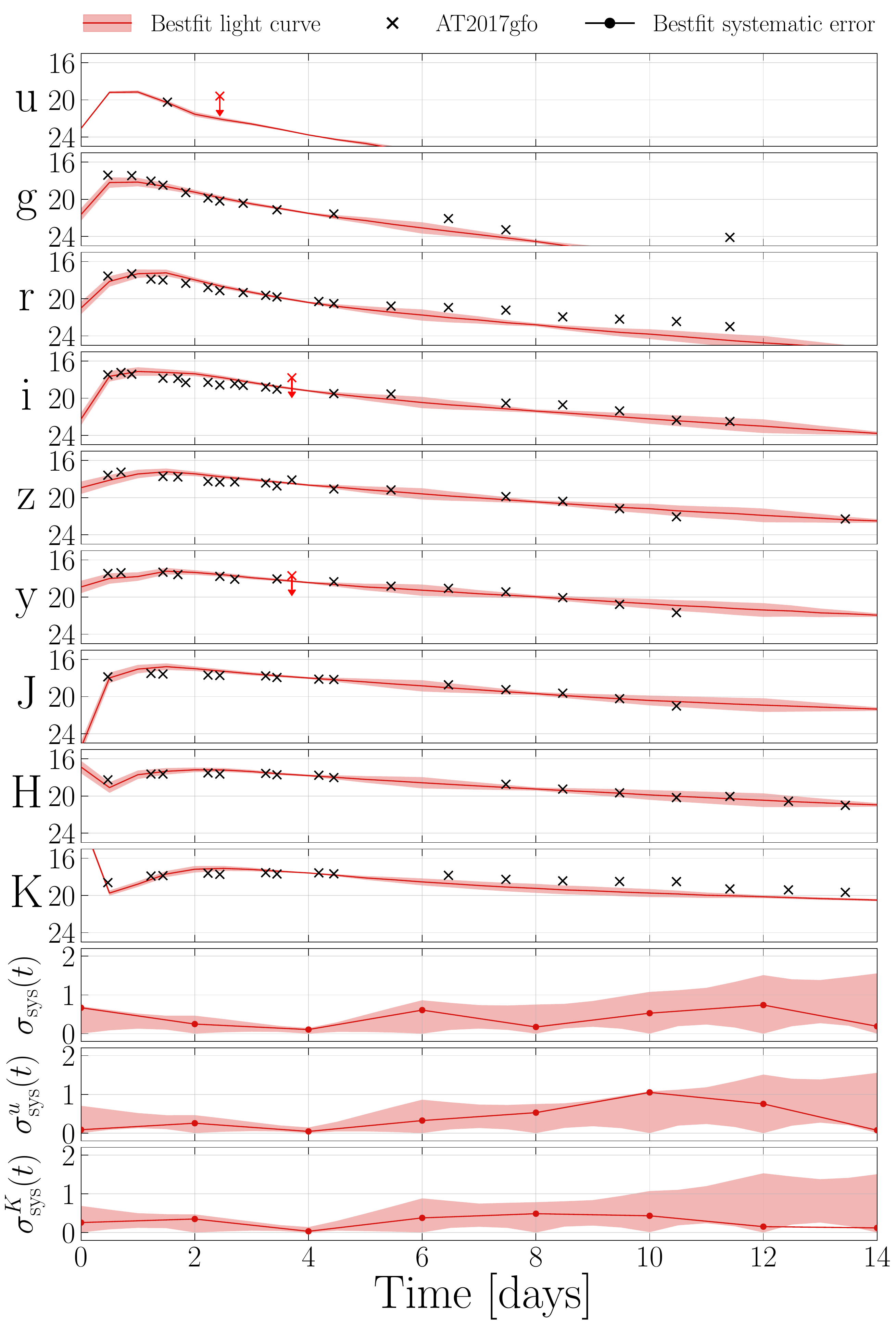}
		\caption{Same as Fig.~\ref{fig:at2017gfo}, for 8 time nodes. $\syserr[]^u(t)$ and $\syserr[]^K(t)$ represents systematic uncertainty for independently sampled $u$ and $K$ band, $\syserr(t)$ represents the systematic uncertainty for all other bands sampled together.}
		\label{fig:at2017gfo_freq}
	\end{center}
\end{figure}

\section{Analyzing AT2017\MakeLowercase{gfo}}
\label{sec:analysis}

\subsection{Time-dependent Uncertainties}
In the previous section, we tested our uncertainty quantification with injections, which confirmed the model's overall robustness but also showed the importance of using physically more complete models, e.g., non-spherical symmetric ones, to interpret observational data. 
To follow up on our tests, we will use the time-dependent uncertainty method to analyze real observational data, focusing on the observed kilonova AT2017gfo. Given the better performance of the \bunine\ model during our validation tests, we will focus purely on this model. 

We will further use this investigation to test the influence of the number of time nodes on the obtained posteriors. For this reason, we performed parameter estimation using seven different numbers of time nodes $N \in \{4,6,8,10,12,14,16\}$. For comparison, we also study AT2017gfo with a constant 1-magnitude error, and a free but time- and filter-independent systematic error, $\syserr \ = \ \mathcal{U}(0,2)$. The values are reported in the first and second rows of Tab.~\ref{tab:at2017gfo}, respectively.

\begin{figure*}[t]
	\centering
	\includegraphics[width=0.815\textwidth]{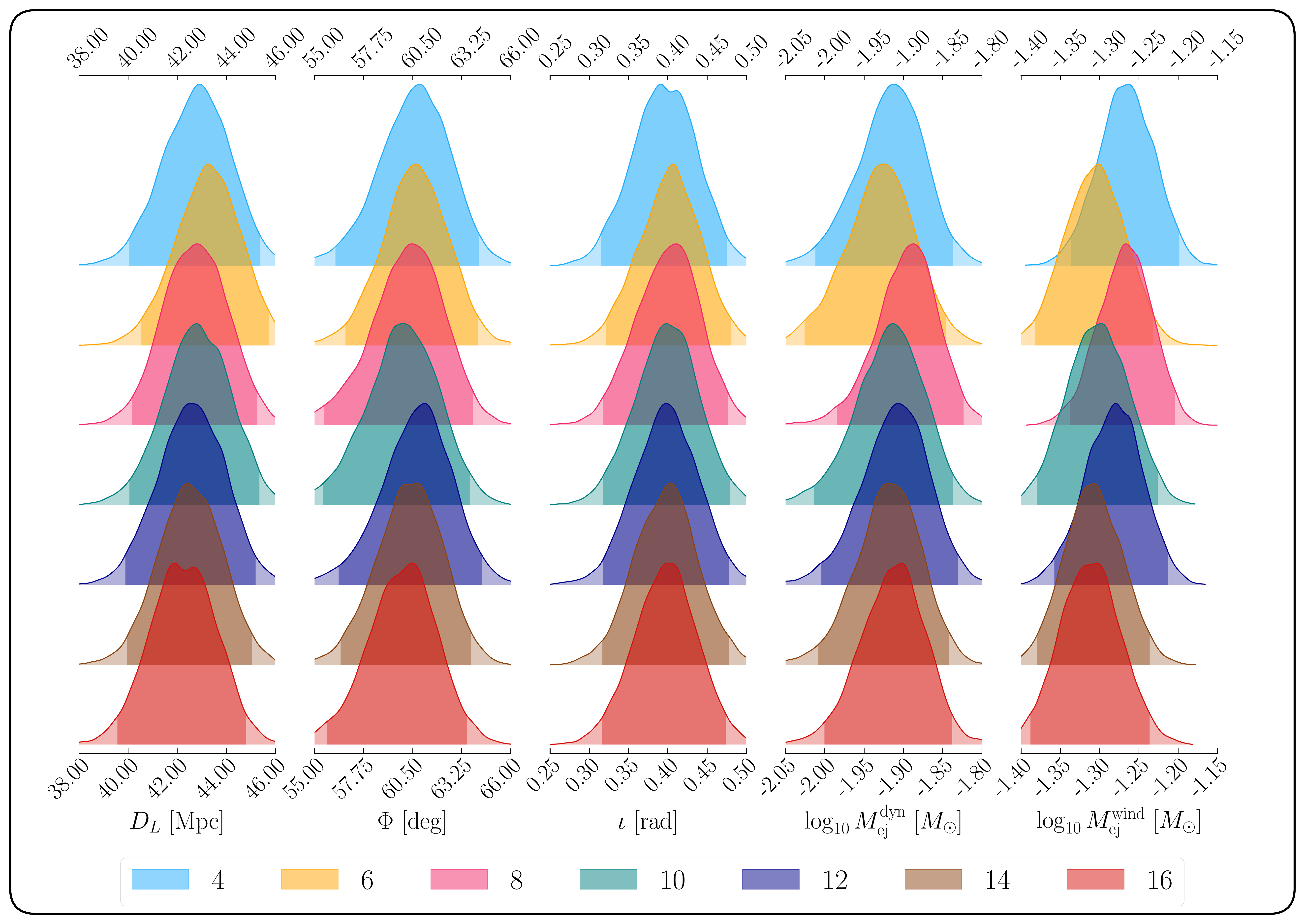}
	\caption{Posterior distribution of the recovered source parameters when analyzing AT2017gfo with the \bunine\ model. Different colors represent different employed number of time nodes.}
	\label{fig:ridge_plot}
\end{figure*}

\begin{table*}[t]
	\centering
	\renewcommand{\arraystretch}{1.25}
 \begin{tabular}{cccccccccc}

		Parameter                                             &
		$D_L \ [\mathrm{Mpc}]$                                &
		$\Phi \ [\mathrm{deg}]$                               &
		$\iota \ [\mathrm{rad}]$                              &
		$\log_{10}M^{\mathrm{dyn}}_{\mathrm{ej}} \ [M_\odot]$ &
		$\log_{10}M^{\mathrm{wind}}_{\mathrm{ej}} \ [M_\odot]$ &
        $\ln \mathcal{L}^{\mathrm{max}}$                       & $\ln \mathcal{L}^{\mathrm{max}}_{\rm ref}$             & 
        Runtime    \\ \hline\hline

		\diagbox[innerwidth=1.5cm]{Nodes}{Prior}
		                                                      & $\mathcal{N}(40,1.89)$    &
		$\mathcal{U}(15,75)$                                  &
		$\mathcal{N}(0.37, 0.04)$                             &
		$\mathcal{U}(-3,-1)$                                  &
		$\mathcal{U}(-3,-0.5)$                                &                           -- & -- & --                                                                                                                \\\hline

-- & ${40.33}_{-2.48}^{+2.76}$ &
${60.53}_{-4.88}^{+4.77}$ &
${0.40}_{-0.09}^{+0.07}$ &
${-1.97}_{-0.07}^{+0.09}$ &
${-1.27}_{-0.05}^{+0.06}$  & $-145.162$ &  ${-51.648}$ & 05 \rm m 57 \rm s \\

 -- & ${40.53}_{-2.85}^{+3.05}$ &
${60.65}_{-8.99}^{+7.19}$ &
${0.37}_{-0.08}^{+0.08}$ &
${-1.96}_{-0.13}^{+0.13}$ &
${-1.26}_{-0.08}^{+0.09}$ & ${-119.275}$ &  ${-25.762}$ & 11 \rm m 16 \rm s \\ \hline

		4                                                     & ${42.83}_{-2.78}^{+2.53}$ & ${60.65}_{-4.48}^{+3.57}$ & ${0.40}_{-0.08}^{+0.08}$ & ${-1.91}_{-0.10}^{+0.08}$ & ${-1.27}_{-0.07}^{+0.07}$ &$-98.971$ & $-5.457$ & 20 \rm m 15 \rm s  \\
		6                                                     & ${43.28}_{-2.75}^{+2.46}$ & ${60.64}_{-3.92}^{+3.50}$ & ${0.40}_{-0.08}^{+0.08}$ & ${-1.93}_{-0.10}^{+0.08}$ & ${-1.31}_{-0.08}^{+0.07}$ & $-97.729$ & $-4.215$ & 33 \rm m 47 \rm s\\
		8                                                     & ${42.75}_{-2.60}^{+2.52}$ & ${60.30}_{-4.76}^{+3.57}$ & ${0.40}_{-0.08}^{+0.07}$ & ${-1.89}_{-0.09}^{+0.07}$ & ${-1.27}_{-0.07}^{+0.06}$ & $-93.514$ & ref. & 46 \rm m 52 \rm s \\
		10                                                    & ${42.83}_{-2.77}^{+2.53}$ & ${59.97}_{-4.52}^{+3.74}$ & ${0.40}_{-0.08}^{+0.08}$ & ${-1.92}_{-0.10}^{+0.08}$ & ${-1.30}_{-0.08}^{+0.08}$ & $-93.630$ & $-0.116$ & 1 \rm h 02 \rm m 48 \rm s\\
		12                                                    & ${42.61}_{-2.72}^{+2.59}$ & ${60.81}_{-4.47}^{+3.58}$ & ${0.40}_{-0.08}^{+0.08}$ & ${-1.91}_{-0.10}^{+0.08}$ & ${-1.28}_{-0.08}^{+0.07}$& $-92.676$ & $0.838$ & 1 \rm h 06 \rm m 50 \rm s \\
		14                                                    & ${42.54}_{-2.59}^{+2.52}$ & ${60.36}_{-3.91}^{+3.40}$ & ${0.40}_{-0.08}^{+0.08}$ & ${-1.92}_{-0.09}^{+0.08}$ & ${-1.31}_{-0.07}^{+0.07}$& $-92.530$ & $0.984$ & 1 \rm h 40 \rm m 02 \rm s \\
		16                                                    & ${42.25}_{-2.70}^{+2.56}$ & ${59.98}_{-4.31}^{+3.58}$ & ${0.40}_{-0.08}^{+0.08}$ & ${-1.91}_{-0.09}^{+0.07}$ & ${-1.31}_{-0.08}^{+0.08}$ & $-92.382$ & $1.132$ &1 \rm h 46 \rm m 40 \rm s\\ \hline
        8 & ${42.56}_{-2.32}^{+2.45}$ &
            ${60.87}_{-3.60}^{+3.27}$ &
            ${0.39}_{-0.07}^{+0.08}$ &
            ${-1.91}_{-0.08}^{+0.07}$ &
            ${-1.29}_{-0.07}^{+0.06}$ & $-94.915$ & $-1.401$ & 56 \rm m 40 \rm s\\ \hline\hline
	\end{tabular}
	\caption{Posterior values with $2\sigma$ credibility, maximum log-likelihoods ($\ln \mathcal{L}^{\mathrm{max}}$), and maximum log-likelihood ratios ($\ln \mathcal{L}^{\mathrm{max}}_{\rm ref}$) values for AT2017gfo  with different systematic configuration.
    \textit{First row: } Values with a constant systematic error of $1~\rm mag$.
    \textit{Second row: } Values with a free, but time- and filter-independent systematic error with a prior of $\mathcal{U}(0,2)$.
    \textit{Last row:} Values with $u$ and $K$ band systematic uncertainties sampled independently.}
	\label{tab:at2017gfo}
\end{table*}

We present the best-fit light curves for the analysis employing different number of time nodes in Fig.~\ref{fig:at2017gfo}; cf.~Fig.~\ref{fig:at2017gfo_freq} for a corresponding analysis incorporating filter-dependent systematic uncertainties.
The posteriors are shown in Fig.~\ref{fig:ridge_plot} and summarized in Tab.~\ref{tab:at2017gfo}. We find that the number of time nodes has only a small influence on the recovered source parameters and that the obtained posteriors agree with their uncertainties. In addition, we can see that the systematic uncertainty (bottom panel of Fig.~\ref{fig:at2017gfo}) is largest for the early times $<1\rm\,day$ and late times $> 5 \rm \,days$. We suggest that this time dependence is caused by (i) the sparseness of observational data in the early times (indicating the need for quick follow-up observations, e.g,~\cite{Shvartzvald:2023ofi}) and (ii) model limitations of \bunine. 
Regarding the latter, higher systematic uncertainties are expected from \texttt{POSSIS} both at early times ($\lesssim 1$ day), due to the implemented opacities being computed only for low-ionization states \cite{Tanaka:2019iqp,Bulla:2022mwo}, and at late times ($\gtrsim5$ days) when the assumed local thermodynamic equilibrium is likely to break down \cite{Pognan:2021wpy,Pognan:2022pix}.

While the posteriors are consistent between different configurations, one needs to gauge which one is sufficiently flexible for representing the underlying systematics. In the context of Bayesian statistics, the most straightforward choice is using the Bayes factors. However, in our use case, the systematic parameters do not represent any physical information; they only represent artificial degrees of freedom. Therefore, to exclude Occam's razor from our decision-making, we compare the maximum log-likelihood between them; cf.~Tab.~\ref{tab:at2017gfo}.

As expected, the most flexible configuration with 16 time nodes achieves the highest maximum log-likelihood. However, due to the marginal increase in log-likelihood and the substantial rise in runtime beyond the eight time nodes configuration, we consider the latter as our reference. For runtime comparison, we use 10 cores on an Intel Xeon Platinum 8270 CPU for each run.



\subsection{Time- and filter-dependent Uncertainties}
\label{ssec:time_filter_uncer}

For comparison, we also apply our method, employing time- and filter-dependent uncertainties to analyze AT2017gfo. As for our validation tests, we group the individual bands such that the $u$ and $K$ bands have individual uncertainties while the other bands are grouped together. We employed eight-time nodes for this analysis. The best-fit lightcurves are shown in Fig.~\ref{fig:at2017gfo_freq}, and the recovered source parameters are summarized in the bottom row of Tab.~\ref{tab:at2017gfo}. The uncertainty in the $u$ band is on the lower side $< 1$ mag, until 8 days despite having only one detection point, and then increases after 8 days as the model cannot perform well due to lack of data. In the $K$ band, where we have a large number of detections, the uncertainty is well constrained throughout, $< 0.5$ mag. The rest of the bands show a similar trend where the uncertainty is again on the lower side, $< 1$ mag. Using filter-dependent systematic uncertainty the maximum log-likelihood decreases marginally; however, the corresponding light curve and associated systematic uncertainty is able to capture the overall trend of the kilonova.\\

\section{Conclusions}
\label{sec:conclusion}
In this work, we have presented a novel approach to quantify the systematic uncertainties in kilonova modeling using time-dependent and filter-dependent interpolation schemes. Our methodology allows for more robust parameter estimation by capturing the non-stationary behavior of systematic errors intrinsic to the underlying models.

Through a series of injection-recovery tests using synthetic light curves from the \katwo\ and \bunine\ models, we have validated the effectiveness of our approach. These tests demonstrate that our interpolation schemes can, in most cases, successfully recover injected parameters within credible intervals.

Applying our methodology to the event AT2017gfo, we performed parameter estimation using different numbers of time nodes. While the best-fit light curves for all runs visually fit the observed data well, a maximum log-likelihood marginal increase criterion favors the model with eight-time nodes as the optimal choice. 

Our work highlights the importance of properly accounting for systematic uncertainties in kilonova modeling, as they can substantially influence the inferred parameters and their uncertainties. By introducing time and filter dependence in treating systematic errors, we provide a more nuanced approach that can be adapted to various electromagnetic models and transient phenomena parameter estimation.\\

\acknowledgements

S.J., T.W., P.T.H.P., and T.D. acknowledge support from the Daimler and Benz Foundation for the project ``NUMANJI''. T.D. acknowledges support from the European Union (ERC, SMArt, 101076369). Views and opinions expressed are those of the authors only and do not necessarily reflect those of the European Union or the European Research Council. Neither the European Union nor the granting authority can be held responsible for them. T.W. and P.T.H.P. are supported by the research programme of the Netherlands Organisation for Scientific Research (NWO). M.B. acknowledges the Department of Physics and Earth Science of the University of Ferrara for the financial support through the FIRD 2024 grant. M.W.C. acknowledges support from the National Science Foundation with grant numbers PHY-2308862 and PHY-2117997. The research leading to these results has received funding from the European Union’s Horizon 2020 Programme under the AHEAD2020 project (grant agreement n. 871158).

\bibliography{refs}{}
\bibliographystyle{apsrev4-1}

\end{document}